\documentstyle[12pt,twoside]{article}
\begin{document}
\title{TROUBLES WITH QUANTUM ANISOTROPIC COSMOLOGICAL MODELS:
LOSS OF UNITARITY}

\author{F.G. Alvarenga\thanks{%
e-mail: flavio@cce.ufes.br}, A.B. Batista\thanks{%
e-mail: brasil@cce.ufes.br},
J. C. Fabris\thanks{%
e-mail: fabris@cce.ufes.br}, S.V.B. Gon\c{c}alves\thanks{%
e-mail: sergio@cce.ufes.br}\\
Departamento de F\'{\i}sica, Universidade Federal do Esp\'{\i}rito Santo, Brazil}
\maketitle
\begin{abstract}
The anisotropic Bianchi I cosmological model coupled with perfect fluid is
quantized in the minisuperspace. The perfect fluid is described by using the
Schutz formalism which allows to attribute dynamical degrees of freedom to
matter. A Schr\"odinger-type equation is obtained where the matter variables play
the role of time. However, the signature of the kinetic term is hyperbolic. This
Schr\"odinger-like equation is solved and a wave packet is constructed. The norm
of the resulting wave function comes out to be time dependent, indicating the loss
of unitarity in this model. The loss of unitarity is due to the fact that the
effective Hamiltonian is hermitian but not self-adjoint.
The expectation value and the bohmian trajectories are
evaluated leading to different cosmological scenarios, what is a consequence of the
absence of a unitary quantum structure. The consistency of this quantum model is
discussed as well as the generality of the absence of unitarity in anisotropic quantum
models.

\vspace{0.7cm}

PACS number(s): 04.20.Cv., 04.20.Me
\end{abstract}

\section{Introduction}

One of the main hopes regarding quantum cosmology is the possibility to obtain
the initial conditions that will determine the ulterior evolution of the Universe
when its classical regime is reached \cite{halliwell,adm}.
The task of obtaining a quantum cosmological scenario faces many difficulties, one
of them being the absence of a natural time variable, since the general relativity action
leads to a constrained system which is invariant under time reparametrization.
There are many attempts to recover the notion of time in quantum cosmology \cite{isham}.
For example, a time coordinate may be identified with the space volume, which is
a growing function in an expanding universe. But, all these attempts have revealed of
limited applications until now, and the problem of time in quantum cosmology remains
an unsolved puzzle.
\par
Another suggestion to incorporate a time variable in quantum cosmology is through matter
fields. This proposal has been extensivelly discussed in reference \cite{kuchar}.
It has been shown that a typical ordinary quantum mechanical structure can be built up:
a Hilbert space, with an inner product, as well as sets of physical observables, may be
identified. All analysis performed in reference \cite{kuchar} was made keeping the functional character
of the Wheeler-DeWitt equation. A simpler consideration in the same sense was made
in references \cite{rubakov,gotay,nivaldo,flavio1,nelson1,nivaldo1,brasil1}, where the matter
fluids were introduced with the aid of the Schutz's variables \cite{schutz1,schutz2},
the Wheeler-DeWitt equation being written in the minisuperspace. The employement
of the Schutz's variables permits again to identify the matter fields with time, since
the associated momentum appears linearly in the Lagrangian; the restriction to minisuperspace
has the advantage of allowing an explicit intregration of the resulting Schr\"ondinger-like
equation.
\par
Although of phenomenological nature, these quantum cosmological perfect fluid models in
the minisuperspace are a very good laboratory in order to verify the consistency of constructing
quantum cosmological models where the time variable is identified with the matter fields.
Since a Schr\"ondinger-like equation is obtained, all the machinery of ordinary quantum
mechanics can be employed. In the references \cite{rubakov,gotay,nivaldo,flavio1,nelson1,nivaldo1,brasil1}, this has been done
for isotropic universes in many different situations, connected mainly to
the nature of the matter content. In order to keep the effective Hamiltonian hermitian, the
inner product in the Hilbert space has acquired an aditional factor. Boundary conditions
on the wave functions were imposed, assuring the hermiticity (and, at the same time, the self-adjointness) of the
effective Hamiltonian operator. Wave packets were constructed from
which specific predictions were obtained by computing the expectation value of the
observables (in this case, the scale factor) or by evaluating the bohmian trajectories.
Since the modulus of the wave function integrated in all space is time independent,
both results agree. The main conclusion of those works is that the quantum model predicts
a singularity-free universe which exhibits a bounce approaching the classical behaviour
asymptotically.
\par
In the present work we will attempt to extend this analysis to anisotropic cosmological
models, specifically to Bianchi I models. The minisuperspace approach will be used, as
well as the description of the matter fields through the Schutz's formalism. The initial
aim is to verify if quantum effects may suppress the anisotropies in the same way they
have suppressed the initial singularity in the isotropic case. In doing this analysis,
an unexpected feature of anisotropic quantum cosmological model appears: the norm of
the wave function
comes out to be time dependent. Hence, the quantum model is non-unitary. This leads immediately
to the question if this is a real quantum system. With a non-unitary theory a probabilistic
interpretation can not be implemented, since the norm of the wave function is not a conserved
quantity. At the same time, the ontological interpretation of Bohm-de Broglie becomes
doubtfull, since bohmian trajectories are not conserved also. In both cases, some
kind of "creation" must be admitted in order to have some quantum interpretation of
the results. If an anisotropic quantum cosmological perfect fluid model is a legitimate
quantizable system, some fundamental changes in quantum mechanical interpretation must
be implemented.
\par
The reason for this loss of unitarity when a simple extension from isotropic to anisotropic
universes is made, is due, in our opinion, to the fact that the kinetic term of the effective Hamiltonian
is not positive definite and, at the same time, due to the measure in the
original gravitational action. The hyperbolic signature of the
kinetic term implies that there is a whole line in the phase space where
the momenta are not zero, and yet the energy is zero. Hence, the wave function need not
to be zero at infinity along this line, and this implies that the probability current is
non-zero at least at some points at infinity. Hence the time derivative of the
norm of the wave function is not zero anymore.
It is important to notice that the effective Hamiltonian is hermitian. The loss
of unitarity comes from the fact that this effective Hamiltonian is not self-adjoint and it
does not admit any self-adjoint extension.
\par
This is not an artifact of the construction of the wave packet, but a general
feature of anisotropic quantum models which lead to an hyperbolic signature to the
kinetic term of the Hamiltonian function with an unusual measure. This suspiction is supported by the fact that
when the kinetic term is made elliptic by force, the norm of the wave function becomes
time independent and normal quantum framework is restablished. But, the hyperbolicity of
the Hamiltonian is not the only reason for the lack of unitarity: the specific measure
in the action, due to the determinant of the metric, is also essential in the sense that
its suppression by force restore also the unitarity independently of the signature of the kinetic term.
Hence, the loss of unitarity is a direct consequence of a gravitational
system (which leads to a non conventional measure in the inner product) which exhibits anisotropies (which leads to a hyperbolic kinetic term).
\par
As it will be verified, the loss of unitarity leads to a fundamental discrepancy between the
many-worlds intepretation (based on the Copenhagen interpretation) \cite{tipler}
and the Bohm-de Broglie
interpretation \cite{holland,nelson} of quantum mechanics. In both cases the universe exhibit a
bounce. But the computation of the
expectation value for the metric functions reveals a universe always isotropic; on the
other hand, the
bohmian trajectories reveal a universe where anisotropies are present near the bounce
disappearing asymptotically. It is important to stress that such quantum Bianchi I model
has been extensivelly studied in the literature \cite{lidsey,gurovich,nelson2,velasco,hawking}. But, in all previous work no matter
field has been used, and hence no explicit time coordinate has been identified. For this
reason, in our opinion, the loss of unitarity has not been remarked before.
\par
This article is organized as follows. In the next section, we construct the Wheeler-DeWitt equation for the anisotropic perfect fluid model and we determine the wave function by using the separation of variables method.
In section 3, a wave packet is constructed and its norm is shown to be time dependent. The
reason for this unexpected result is discussed. In section 4, the expectation value of the scale factor and the bohmian trajectories are obtained. The discrepancy between them
are settled out. In section 5 we discuss the results and present our conclusions.
In the appendix we show how the equivalence between many-worlds and dBB interpretation
disappears due to the absence of unitarity in a quantum model. 

\section{Wheeler-DeWitt equation for an anisotropic perfect fluid model}

Our starting point is the action of gravity coupled to a perfect fluid in the Schutz's formalism:
\begin{equation}
\label{action}
{\cal A} = \int_Md^4x\sqrt{-g}R + 2\int_{\partial M}d^3x\sqrt{h}h_{ab}K^{ab}
+ \int_Md^4x\sqrt{-g}p 
\end{equation}
where $K^{ab}$ is the extrinsic curvature, and $h_{ab}$ is the induced
metric over the three-dimensional spatial hypersurface, which is
the boundary $\partial M$ of the four dimensional manifold $M$; the
factor $16\pi G$ is made equal to one.
The first two terms were first obtained in reference \cite{adm};
the last term of (\ref{action}) represents the matter contribution to
the total action in the Schutz's formalism for perfect fluids, $p$ being the pressure, which is linked to the energy density by the equation of state $p = \alpha\rho$. In the
Schutz's formalism \cite{schutz1,schutz2},
the four-velocity is expressed in terms of five potentials $\epsilon$,
$\zeta$, $\beta$, $\theta$ and $S$:
\begin{equation}
U_\nu = \frac{1}{\mu}(\epsilon_{,\nu} + \zeta\beta_{,\nu} +
\theta S_{,\nu})
\end{equation}

\noindent where $\mu$ is the specific enthalpy. The variable $S$ is the specific
entropy, while the potentials $\zeta$ and $\beta$ are connected with
rotation and are absent for FRW's type models. The variables $\epsilon$ and
$\theta$ have no clear physical meaning.
The four velocity is subject to the condition
\begin{equation}
U^\nu U_\nu = 1 \quad .
\end{equation}

\par
The metric describing a Bianchi I anisotropic model is given by
\begin{equation}
ds^2 = N^2dt^2 - \biggr(X(t)^2dx^2 + Y(t)^2dy^2 + Z(t)^2dz^2\biggl) \quad .
\end{equation}
In this expression, $N(t)$ is the lapse function.
Using the constraints for the fluid, and after some thermodynamical considerations,
the final reduced action, where surface terms were discarded,
takes the form
\begin{eqnarray}
{\cal A} = \int dt\biggr[-\frac{2}{N}\biggr(\dot X\dot YZ + \dot X\dot ZY +
\dot Y\dot ZX\biggl) \nonumber\\
+ N^{-1/\alpha}
(XYZ)\frac{\alpha}{(\alpha + 1)^{1/\alpha + 1}}(\dot\epsilon +
\theta\dot S)^{1/\alpha + 1}\exp\biggr(- \frac{S}{\alpha}\biggl)
\biggl] \quad .
\end{eqnarray}
\par
At this point, is more suitable to redefine the metric coefficients as
\begin{equation}
X(t) = e^{\beta_0 + \beta_+ + \sqrt{3}\beta_-}\quad ,\quad Y(t) = e^{\beta_0 + \beta_+ - \sqrt{3}\beta_-}\quad ,
\quad Z(t) = e^{\beta_0 - 2\beta_+} \quad . 
\end{equation}
Using these new variables, the action may be simplified further, leading
to the gravitational Lagrangian density
\begin{equation}
L_G = -6\frac{e^{3\beta_0}}{N}\{\dot\beta_0^2 - \dot\beta_+^2 - \dot\beta_-^2\} \quad .
\end{equation}

From this expression, we can evaluate the conjugate momenta:
\begin{equation}
p_0 = - 12\frac{e^{3\beta_0}}{N}\dot\beta_0 \quad , \quad p_+ = 12\frac{e^{3\beta_0}}{N}\dot\beta_+ \quad ,
\quad
p_- = 12\frac{e^{3\beta_0}}{N}\dot\beta_- \quad .
\end{equation}

The matter sector may be recast in a more suitable form through the canonical transformations
\begin{equation}
T = p_Se^{-S}p_\epsilon^{-(\alpha + 1)} \quad , \quad 
p_T = p_\epsilon^{\alpha + 1}e^S \quad , \quad
\bar\epsilon = \epsilon - (\alpha + 1)\frac{p_S}{p_\epsilon} \quad ,
\quad \bar p_\epsilon = p_\epsilon \quad .
\end{equation}

The final expression for the total Hamiltonian is
\begin{equation}
\label{hamilton}
H = Ne^{-3\beta_0}\biggr\{- \frac{1}{24}(p_0^2 - p_+^2 - p_-^2) + e^{3(1-\alpha)\beta_0}p_T\biggl\} \quad .
\end{equation}
The fundamental aspect of the Hamiltonian (\ref{hamilton}) to be remarked is the
hyperbolic signature of the kinetic term.
\par
The lapse function $N$ plays the role of a Lagrangian multiplier in (\ref{hamilton}).
It leads to the constraint
\begin{equation}
H = 0 \quad .
\end{equation}
The quantization procedure consists in considering the Hamiltonian as
an operator which is applied on a wave function
\begin{equation}
\hat H\psi = 0
\end{equation}
taking at the same time the momenta as operators (we use natural units where $\hbar = 1$):
\begin{equation}
\hat p_i = -i\frac{\partial}{\partial\beta_i} \quad .
\end{equation}
Since the momentum associated to the matter degrees of freedom appears linearly in the Hamiltonian,
we can identify it with a time coordinate
\begin{equation}
\hat p_T = i\frac{\partial}{\partial T} \quad .
\end{equation}
Due to the canonical transformations employed before,
this new time is related to the cosmic time $t$ by
$dt = e^{3\alpha\beta_0}dT$.
In this way, we end up with the Wheeler-DeWitt equation, in the minisuperspace, for an anisotropic Universe
filled with a perfect fluid:
\begin{equation}
\label{wdwe}
\biggr(\frac{\partial^2}{\partial\beta^2_0} - \frac{\partial^2}{\partial\beta^2_+} -
\frac{\partial^2}{\partial\beta^2_-}\biggl)\psi = - 24ie^{3(1 - \alpha)\beta_0}\frac{\partial\psi}{\partial T} \quad .
\end{equation}
\par
The wave function $\Psi$ must obey the following boundary conditions:
\begin{equation}
\Psi'|_{\beta_i \rightarrow \pm \infty} = \kappa\Psi|_{\beta_i \rightarrow \pm \infty} \quad ,
\end{equation}
with $\kappa \in (- \infty,\infty]$, $\beta_i$ denoting the dynamical variables. These boundary conditions are established by requiring
that the Hamiltonian be hermitian. For $\kappa = 0$ and $\infty$, the boundary conditions are $\Psi'|_{\beta_i \rightarrow \pm \infty} = 0$ and $\Psi|_{\beta_i \rightarrow \pm \infty} = 0$, respectively. As it will be seen later, in spite of being hermitian, the effective
Hamiltonian is not self-adjoint and does not admit any self-adjoint extension.
This will lead to the loss of unitarity.
A more detailed discussion on the
self-adjoint properties of the operators in quantum cosmology with perfect fluid,
in a situation very close to the present one, is given in references \cite{nivaldo,brasil1}.
A rigorous mathematical discussion is given in reference \cite{farhi}.
\par
Now, our goal is to solve (\ref{wdwe}) and to construct the corresponding wave packet.
To do so, we use the separation of variable's method.
First, we write the wave function as
\begin{equation}
\psi(\beta_0,\beta_+,\beta_-,T) = \phi(\beta_0,\beta_+,\beta_-) e^{-iET} \quad ,
\end{equation}
leading to the equation
\begin{equation}
\biggr(\frac{\partial^2}{\partial\beta^2_0} - \frac{\partial^2}{\partial\beta^2_+} -
\frac{\partial^2}{\partial\beta^2_-}\biggl)\phi = - 24Ee^{3(1 - \alpha)\beta_0}\phi \quad .
\end{equation}
The function $\phi$ is then
written as
\begin{equation}
\phi(\beta_0,\beta_+,\beta_-) = \Upsilon_0(\beta_0)\Upsilon_+(\beta_+)\Upsilon_-(\beta_-) \quad ,
\end{equation}
leading to the equation
\begin{equation}
\frac{\partial^2_0\Upsilon_0}{\Upsilon_0} + 24Ee^{3(1 - \alpha)\beta_0}
- \frac{\partial^2_+\Upsilon_+}{\Upsilon_+} - \frac{\partial^2_-\Upsilon_-}{\Upsilon_-} = 0
\end{equation}
where we have simplified in an obvious way the notation for the partial derivatives.
The solutions for the functions $\Upsilon_\pm$ are
\begin{equation}
\Upsilon_\pm = C_\pm e^{ik_\pm\beta_\pm} \quad ,
\end{equation}
where $C_\pm$ are constants and $k_\pm$ are the separation parameters. These separation parameters
must be real otherwise the wave function is not normalizable.
\par
The equation determining the behaviour of $\Upsilon_0$ takes then the form,
\begin{equation}
\label{de}
{\Upsilon_0}'' + \biggr(24Ee^{3(1-\alpha)\beta_0} + (k_+^2 + k_-^2)\biggl)\Upsilon_0 = 0 \quad,
\end{equation}
the primes meaning derivatives with respect to $\beta _0$. It is easily to see that the parameter
$E$ must be positive.
The previous equation can be solved through the redefinitions
\begin{equation}
a = e^{\beta_0} \quad , \quad y = a^r \quad , \quad r = \frac{3}{2}(1 - \alpha) \quad .
\end{equation}
after what (\ref{de}) takes the form of a Bessel's equation:
\begin{equation}
{\ddot\Upsilon}_0 + \frac{{\dot\Upsilon}_0}{y} + \biggr(\frac{24E}{r^2} + \frac{k^2}{r^2}\frac{1}{y^2}\biggl)\Upsilon_0 = 0
\end{equation}
where $k^2 = k^2_+ + k^2_-$ and the dots are derivatives with respect to $y$.
The solution is
\begin{equation}
\Upsilon_0 = C_1J_\nu\biggr(\frac{\sqrt{24E}}{r}a^r\biggl) + C_2J_{-\nu}\biggr(\frac{\sqrt{24E}}{r}a^r\biggl) \quad ,
\end{equation}
with $\nu = ik/r$, $C_{1,2}$ being integration constants.
\par
The final expression for the wave function is then
\begin{equation}
\label{sol}
\Psi = e^{i(k_+\beta_+ + k_-\beta_-)}\biggr[\bar C_1J_\nu\biggr(\frac{\sqrt{24E}}{r}a^r\biggl) +
\bar C_2J_{-\nu}\biggr(\frac{\sqrt{24E}}{r}a^r\biggl)\biggl]e^{-iET}
\end{equation}
where $\bar C_{1,2}$ are combinations of the preceding integration constants.

\section{The wave packet: Loss of unitarity}

We want now to construct a superposition of the solutions (\ref{sol}), generating a regular wave packet.
In principle, this can be achieved by considering the integration constants as gaussian functions of the
parameters $k_\pm$ and $E$. The general case constitutes a hard problem from the
technical point of view. We may consider, for simplicity, the final wave function independent of one of the variables $\beta_\pm$, which amounts
to fix one the corresponding parameters $k_+$ or $k_-$ equal to zero. 
From here on we will consider
$k_- = 0$. Notice that the final results would
be the same if we had imposed $k_+ = 0$ and $k_- \neq 0$. Hence, even if the anisotropic models are not
analyzed in all their generality, a large class of them is covered in what follows.
\par
Fixing $k_- = 0$, the wave packet is given by
\begin{equation}
\label{eo}
\Psi = \int e^{ik_+\beta_+}\biggr\{\bar C_1J_\nu\biggr(\frac{\sqrt{24E}}{r}a^r\biggl) +
\bar C_2J_{-\nu}\biggr(\frac{\sqrt{24E}}{r}a^r\biggl)\biggl\}e^{-iET}dk_+dE \quad .
\end{equation}
In principle, in the expression for $\nu$ it appears the modulus of $k_+$ while in the first exponential in (\ref{eo})
we have $- \infty < k_+ < + \infty$.
We will consider a superposition of both Bessel's functions in such a way that the expression
for the wave packet may be written as
\begin{equation}
\Psi = \int_{-\infty}^{+\infty}\int_0^\infty A(k_+,q)e^{ik_+\beta_+}J_\nu\biggr(qa^r\biggl)e^{-iq^2T}dk_+dq \quad ,
\end{equation}
with $q = \frac{\sqrt{24E}}{r}$ and
\begin{equation}
\label{sup}
A(k_+,q) = e^{-\gamma k_+^2}q^{\nu + 1}e^{-\lambda q^2} \quad .
\end{equation}
In this case, the integrals can be explicitly calculated, leading to the wave packet
\begin{equation}
\label{wp}
\Psi = \frac{1}{B}\sqrt{\frac{\pi}{\gamma}} \exp\biggr[- \frac{a^{2r}}{4B} - \frac{(\beta_+  + C(a,B))^2}{4\gamma}\biggl]
\end{equation}
where
\begin{equation}
B = \lambda + isT \quad , \quad C(a,B) = \ln a - \frac{2}{3(1 - \alpha)}\ln 2B \quad , \quad
s = - \frac{3(1 - \alpha)^2}{32} \quad .
\end{equation}
Notice that the wave packet given by (\ref{wp}) is square integrable, and it vanishes in the extremes of the interval of validity of
the variables $a = e^{\beta_0}$ and $\beta_+$, except along the line $\beta_0 = - \beta_+$ where it
takes a constant value, being consequently regular as it is physically required.
The wave packet (\ref{wp}) is indeed a solution of the equation (\ref{wdwe}), as it can
be explicitly verified, and it obeys the boundary conditions fixed before. If we discard
the terms corresponding to the variable $\beta_+$(connected with $k_+$), the wave packet for
the isotropic case \cite{nivaldo1} is reobtained.
\par
The main point to be remarked now is that the norm of (\ref{wp}) is time dependent.
Using the definition $a = e^{\beta_0}$ and integrating in
$\beta_+$ and $a$ we obtain
\begin{equation}
\int_0^\infty\int_{-\infty}^{\infty}a^{2-3\alpha}\Psi^*\Psi\,da\,d\beta_+ 
= \frac{\sqrt{2\gamma\pi}}{3(1 - \alpha)}\frac{2}{\lambda}F(T)\quad .
\end{equation}
where
\begin{equation}
F(T) = \exp{(\frac{C_I^2}{2\gamma})} \quad ,
\end{equation}
and
\begin{eqnarray}
C(a,B) &=& C_R + iC_I \quad ,\nonumber\\
C_R = \ln a - \frac{1}{3(1 - \alpha)}\ln 4B^*B \quad &,&
\quad C_I = \frac{- 2}{3(1 - \alpha)}\arctan(\frac{sT}{\lambda})
\quad .
\end{eqnarray}
The norm of the wave function is time dependent. Hence, the quantum
model is not unitary. 
\par
The absence of unitarity may be understood by inspecting again the
wave packet (\ref{wp}). In fact, this wave packet goes to zero at infinity,
excepted along the line $\beta_0 = - \beta_+$, where it takes a constant value
at infinity. This does not spoil the regularity of the wave packet; in particular,
it remains finite when integrated in all space and
specific boundary conditions are obeyed. But, this leads, at the
same time, to an anomaly in
the infinity boundary. The reason for this anomaly may be understood by analysing again
the Schr\"odinger-like equation (\ref{wdwe}). Notice that, after decomposing it into
stationary states, the energy $E$ is zero along the whole line $\beta_0 = - \beta_+$.
Along this line, the wave function need not to vanish.
\par
It would be expected that a hermitian Hamiltonian operator should always lead to
a unitary quantum system, since the Hamiltonian operator is responsible for the
time evolution of the quantum states. The problem here relies on the fact that, in
spite of being hermitian, the Hamiltonian effective operator
\begin{equation}
\label{eh}
H_{eff} = e^{-3(1 - \alpha)}\biggr\{\partial_0^2 - \partial_+^2 - \partial_-^2\biggl\}
\end{equation}
is not self-adjoint. This
means that $H^\dag = H$ but the domain of $H^\dag$ is not the same as the domain of
$H$, and the conservation of the norm becomes senseless \cite{symon}.
\par
In order to verify if an operator is self-adjoint or not, we must compute the so-called
deficiency indices $n_\pm$ which are the dimensions of the linear independent square integrable
solutions of the indicial equation
\begin{equation}
H\phi = \pm i\phi \quad .
\end{equation}
Using the effective Hamiltonian (\ref{eh}), the solutions of the indicial equations are
\footnote{It must be remarked that in finding the solutions, we supposed the system to
be independent of the variable $\beta_-$ and we perform a plane wave expansion in the
variable $\beta_+$. In this sense, we considered just a one-dimensional system depending
on the variable $\beta_0$ with a parameter $k$. However, the conclusions do not depend
on these considerations and we could consider at least a two-dimensional system by,
for example, performing a gaussian superposition in the parameter $k$.}
\begin{eqnarray}
\phi_+ = c_1J_\nu(y) + c_2J_{-\nu}(y) \quad , \\
\phi_- = c_3K_\nu(y) + c_4I_\nu(y) \quad ,
\end{eqnarray}
where $\nu = ik/r$ and $y = \sqrt{i/r^2}\,a^r$, $a$ and $r$ having the same definitions as
before. It is easy to see that $J_{\pm\nu}(y)$ and $I_\nu(y)$ are not square integrable solutions while $K_\nu(y)$ is. Hence, $n_+ = 0$ and $n_- = 1$ and, as explained in
\cite{symon}, the effective Hamiltonian operator is not self-adjoint and does not
admit any self-adjoint extension. Notice that changing arbitrarily the signature
in (\ref{eh}) or suppressing the unusual measure, the deficiency indices become
$n_+ = n_- = 0$ and the effective Hamiltonian becomes self-adjoint.
\par
It is important to remark also that, due to hyperbolic character of the Hamiltonian,
the energy $E$ may take negative values. Hence, we may expect that this system is
unstable since the energy is not bound below. However, it is possible to consider
a kind of "Dirac sea" hypothesis, with all negative energy state filled, and to take
effectively into account only positive energy states.
\par
To support the idea that this anomaly is due to the hyperbolic signature in the
kinetic term in (\ref{wdwe}), together with the unusual measure, let us change it to an elliptic signature by force.
In doing so, the main change in the wave functions (\ref{wf}) is that the order of the
Bessel function becomes real: $\nu = |k|$. We keep only the Bessel function
of positive order because it does not diverge as $a \rightarrow 0$.
In evaluating the norm of the wave function, we consider the same superposition factor as in (\ref{sup}) and
we first integrate on the parameter $q$, obtaining,
\begin{equation}
\Psi =  \frac{\Psi_0}{B}\exp{\frac{a^{2r}}{4B}}\int_0^\infty\exp{\biggr[-\gamma k^2 +\biggr(C(a,B)+ i\beta_+\biggl)k\biggl]}dk \quad .
\end{equation}
Now we writte
\begin{equation}
N = \int_{-\infty}^{+\infty}\int_0^\infty a^{2 - 3\alpha}\Psi^*\Psi da\,d\beta_+ \quad ,
\end{equation}
and we integrate first in $\beta_+$ and then in $a$.
The unusual measure in the integrals is due to the requirement that the reduced Hamiltonian in (\ref{wdwe}) must
be hermitian \cite{nivaldo1,brasil1}.
The final result
is
\begin{equation}
N = {\Psi'}_0^2\int_0^\infty 2^{k/r}\Gamma(1 + k/r)e^{-2\gamma k^2} \quad,
\end{equation}
where $\Gamma(x)$ is the gamma function, and $\Psi_0'$ is a new constant. The norm
of the wave function is finite and, more important, time independent. The same occurs
if instead the measure is suppressed.

\section{The scenario for the Universe}

The fact that the quantum cosmological perfect fluid model leads to a non-unitary
quantum system implies in principle that no usual quantum interpretation can
be be applied to it, unless we allow creation of universes. In what follows we will
adopt the point of view that this is a legitimate quantum system which ask for a
convenient framework interpretation.
Hence, we will try to extract previsions for the evolution of such a universe
using the many-worlds and dBB interpretations scheme. Of course, these interpretations scheme
must
be enlarged in order to incorporate non-unitary quantum system. It is not sure that
this can be done consistently. However, this
very important conceptual
problem is
outside the purpose of the present work. Here, in particular we will show that
the many-worlds and Bohm-de Broglie interpretations leads to different results.
This is due to the lack of unitarity as it is explained in the appendix.
\par
Before to do this, let us just recall the classical solutions for the Bianchi I cosmological model
with a barotropic perfect fluid described by $p = \alpha\rho$. For the time parametrization
$dt = a^{3\alpha}dT$, $t$ being the cosmic time, the functions $X$, $Y$, and $Z$ admit the solution
\begin{eqnarray}
X(T) &=& e^{\beta_0 + \beta_+ + \sqrt{3}\beta_-}= X_0\biggr(T + c\biggl)^\frac{1 + 2s_1}{3(1 - \alpha)}\biggr(T - c\biggl)^\frac{1 - 2s_1}{3(1 - \alpha)} \quad ,
\\
Y(T) &=& e^{\beta_0 + \beta_+ - \sqrt{3}\beta_-} = Y_0\biggr(T + c\biggl)^\frac{1 + 2s_2}{3(1 - \alpha)}\biggr(T - c\biggl)^\frac{1 - 2s_2}{3(1 - \alpha)} \quad ,
\\
Z(T) &=& e^{\beta_0 - 2\beta_+} = Z_0\biggr(T + c\biggl)^\frac{1 + 2s_3}{3(1 - \alpha)}\biggr(T - c\biggl)^\frac{1 - 2s_3}{3(1 - \alpha)} \quad ,
\end{eqnarray}
where $c$ is constant, and $s_1$, $s_2$ and $s_3$ are parameters such that
\begin{equation}
s_1 + s_2 + s_3 = 0 \quad , \quad s_1^2 + s_2^2 + s_3^2 = 6 \quad .
\end{equation}
Notice that there is an initial singularity, near which the Universe is very anisotropic, becoming isotropic
asymptotically.
\par
Let us return now to the computation of the quantum scenario through the use of the many-worlds and
ontological interpretations of quantum mechanics.

\subsection{Expectation values of the dynamical variables}

Given the wave function $\Psi$, the expectation value of a variable $\beta_i$ is obtained in the usual way:
\begin{equation}
\label{ev}
<\beta_i> = \frac{\int_{-\infty}^{+\infty}\int_{-\infty}^{+\infty}e^{3(1 - \alpha)\beta_0}\Psi^*\beta_i\Psi d\beta_0d\beta_+}{\int_{-\infty}^{+\infty}\int_{-\infty}^{+\infty}e^{3(1 - \alpha)\beta_0}\Psi^*\Psi d\beta_0d\beta_+} \quad .
\end{equation}
For $\beta_i = \beta_0$ in (\ref{ev}) we find for the numerator:
\begin{equation}
\int_0^\infty a^{2-3\alpha}\Psi^*\Psi\ln a\;da\;d\beta_+ = \frac{F(T)\sqrt{2\gamma\pi}}{9(1 - \alpha)^2}
\frac{2}{\lambda}\biggr\{\ln\biggr(\frac{2B^*B}{\lambda}\biggl)
+ n\biggl\} \quad ,
\end{equation}
where we have noted
\begin{equation}
\quad n
= \int_0^\infty \exp(-u)\ln u\,du \sim - 0.577\quad ,
 \quad u = \frac{\lambda}{2B^*B}a^{3(1 - \alpha)}
\quad .
\end{equation}
Hence,
\begin{equation}
<\beta_0> = \frac{1}{3(1 - \alpha)}\biggr\{\ln\biggr(\frac{2\vert B\vert^2}{\lambda}\biggl) + n\biggl\}
\quad .
\end{equation}
This result leads to
\begin{equation}
e^{<\beta_0>} = (XYZ)^{1/3} = a_0\biggr[1 + \frac{s^2T^2}{\lambda^2}\biggl]^{\frac{1}{3(1 - \alpha)}} \quad ,
\end{equation}
where $a_0$ is a constant. This is the same result as in the isotropic
case \cite{nivaldo1}. Consequently, the space volume evolves as in the corresponding isotropic case.
\par
The anisotropies are represented by the function $\beta_+$, whose expectation value will be computed in what
follows.
We will evaluate now the numerator of (\ref{ev}) with $\beta_i = \beta_+$. Integrating in $\beta_+$ and
expressing $\beta_0$ in terms of $a$ as before, we find:
\begin{eqnarray}
\int_{-\infty}^{+\infty}\int_{-\infty}^{+\infty}e^{3(1 - \alpha)}\Psi^*\beta_+\Psi\;d\beta_0\;d\beta_+ =
- \sqrt{\pi}\biggr\{I_1 - \frac{\ln(4B^*B)}{3(1 - \alpha)}I_2\biggl\}\frac{F(T)}{B^*B} \quad , \\
I_1 = \int_0^\infty a^{2-3\alpha}\exp\biggr\{- \lambda\frac{a^{3(1 - \alpha)}}{2B^*B}\biggl\}\ln a\;da
\quad , \quad I_2 = \int_0^\infty a^{2-3\alpha}\exp\biggr\{- \lambda\frac{a^{3(1 - \alpha)}}{2B^*B}\biggl\} da
\end{eqnarray}
The integrals $I_1$ and $I_2$ take the form,
\begin{equation}
I_1 = \frac{1}{9(1 - \alpha)^2}\biggr[\frac{2B^*B}{\lambda}\biggl]\biggr[n + \ln\biggr(\frac{2B^*B}{\lambda}\biggl)\biggl] \quad , \quad
I_2 = \frac{1}{3(1 - \alpha)}\frac{2B^*B}{\lambda} \quad .
\end{equation}
We find finally
\begin{equation}
<\beta_+> = \frac{1}{3(1 - \alpha)}\biggr\{\ln(2\lambda) - n\biggl\} \quad .
\end{equation}
The expectation value of $\beta_+$ does not depend on time. Consequently, the predicted result
for the evolution of the Universe in this case is the same as in the isotropic case: there is no
anisotropy during all the evolution of the Universe. A similar computation shows that
$<\beta_+^2>$ is also time independent. The cosmological scenario is really isotropic.

\subsection{Computation of the bohmian trajectories}

The result found in the last section indicates no trace of the anisotropies existing
in the classical model in the corresponding quantum analysis. 
We will evaluate the bohmian trajectories
which determine the behaviour of a quantum system in the ontological interpretation of quantum mechanics.
\par
In the ontological interpretation of quantum mechanics, the wave function is written as
\begin{equation}
\label{wf}
\Psi = \Sigma\;\exp(i\Theta) \quad ,
\end{equation}
where $\Sigma$ is connected with the amplitude of the wave function, and $\Theta$ to its phase. When (\ref{wf}) is
inserted into the Schr\"odinger's equation, the real and imaginary parts of the resulting expression leads
to the conservation of probability and to a Hamilton-Jacobi's equation supplemented by a term which is identified
as the quantum potential, which leads to the quantum effects distinguishing the quantum trajectories from
the classical ones.
\par
In this formulation of quantum mechanics, the trajectories (which are real trajectories)
corresponding to a dynamical variable $q$ with a conjugate momentum $p_q$ are given by
\begin{equation}
p_q = \frac{\partial \Theta}{\partial q} \quad .
\end{equation}
The ontological formulation of quantum mechanics
leads to a natural identification of a time coordinate, what is very important for
quantum cosmology where
in general there is no explicit time coordinate.
\par
Let us consider the wave function (\ref{wp}). Putting in the form (\ref{wf}), the phase reads,
\begin{equation}
\Theta(\beta_0,\beta_+,T) = - \arctan\biggr(\frac{sT}{\lambda}\biggl) + \frac{sTa^{3(1 - \alpha)}}{4B^*B} -
\frac{C_I}{2\gamma}(\beta_+ + C_R) \quad ,
\end{equation}
where all quantities are defined as before.
The conjugate momenta associated to the dynamical variables $\beta _0$ and $\beta _+$ read
\begin{equation}
p_0 = - 12a^{2 - 3\alpha}\dot a \quad , \quad p_+ = 12a^{3(1 - \alpha)}\dot\beta_+ \quad ,
\end{equation}
where we have explicitly used the time parametrization such that the lapse function is given
by $N = a^{3\alpha}$.
The bohmian trajectories are then given by the expressions
\begin{eqnarray}
\label{bt1}
- 12a^{2 - 3\alpha}\dot a &=& 3(1 - \alpha)\frac{sT}{4B^*B}a^{3(1 - \alpha)} - \frac{C_I}{2\gamma} \quad , \\
\label{bt2}
12a^{3(1 - \alpha)}\dot\beta_+ &=& - \frac{C_I}{2\gamma} \quad ,
\end{eqnarray}
dots representing derivatives with respect to $T$. Combining (\ref{bt1},\ref{bt2}), we find
\begin{equation}
- 12a^{2 - 3\alpha}\dot a = 3(1 - \alpha)\frac{sT}{4B^*B}a^{3(1 - \alpha)} + 12a^{3(1 - \alpha)}\dot\beta_+ \quad .
\end{equation}
This last equation leads after integration to the expression
\begin{equation}
a\,e^{\beta_+} = D\biggr[\lambda^2 + s^2T^2\biggl]^\frac{1}{3(1 - \alpha)} \quad .
\end{equation}
Reinserting the relation in the equations (\ref{bt1},\ref{bt2}) we can obtain the following solutions
to $a$ and $\beta_+$:
\begin{eqnarray}
a &=& \biggr(\frac{-1}{24s\lambda\gamma}\biggl)^\frac{1}{3(1 - \alpha)}\biggr[\lambda^2 + s^2T^2\biggl]^\frac{1}{3(1 - \alpha)}\biggr[\arctan^2\biggr(\frac{sT}{\lambda}\biggl) + E\biggl]^\frac{1}{3(1 - \alpha)} \quad , \\
\beta_+ &=& - \frac{1}{3(1 - \alpha)}\ln\biggr\{\arctan^2\biggr(\frac{sT}{\lambda}\biggl) + E\biggl\} + \ln\biggr\{\biggr[-24s\lambda\gamma\biggl]^\frac{1}{3(1- \alpha)}\biggl\} + \ln D
\quad ,
\end{eqnarray}
where $E$ and $D$ are integration constants. Remember that $s < 0$.
\par
In opposition to the expressions obtained for the expectation values of $\beta_0$ (which is connected to $a$)
and $\beta_+$ in the preceding subsection, the bohmian trajectories predict an anisotropic Universe.
Until this point, this strange discrepancy is not so catrastrophic: in order the bohmian trajectories coincide with
the results for the expectation value for some quantity, the integration constants that appear in the
former must be averaged over an initial distribution given by the modulus of the wave function at $T = 0$.
At $T = 0$, we have
\begin{eqnarray}
a(T=0) &=& \biggr(\frac{- \lambda E}{24s\gamma}\biggl)^\frac{1}{3(1 - \alpha)} \quad , \\
\beta_+(T=0) &=& \ln\biggr\{\biggr[\frac{- 24s\lambda\gamma D^{3(1 - \alpha)}}{E}\biggl]^\frac{1}{3(1 - \alpha)}\biggl\} \quad .
\end{eqnarray}
Hence,
\begin{equation}
{\cal R_0} = \Psi^*\Psi\vert_{T=0} = \frac{\pi}{\lambda^2\gamma}e^{\frac{E}{48s\gamma} - \frac{1}{18(1 - \alpha)^2\gamma}\ln^2\biggr[\frac{D^{3(1 - \alpha)}}{4}\biggl]} \quad .
\end{equation}
\par
For $\beta_0$ and $\beta_+$ the average on the initial conditions leads to the integral expressions
\begin{eqnarray}
\bar \beta_0(T) &=& \int_0^\infty\int_0^\infty e^{3(1 - \alpha)\beta_0^i}{\cal R_0}\;\beta_0(T)\;d\beta_0^i\;d\beta_+^i \quad , \\ 
\bar \beta_+(T) &=& \int_0^\infty\int_0^\infty e^{3(1 - \alpha)\beta_0^i}{\cal R_0}\;\beta_+(T)\;d\beta_0^i\;d\beta_+^i \quad ,
\end{eqnarray}
where $\beta_0^i$ and $\beta_+^i$ denote the initial values of the metric functions.
In the isotropic case \cite{nivaldo1} the expression corresponding to the
above ones leads to a perfect agreement
between many-worlds and dBB interpretations.
These expressions can be recast in the following form:
\begin{eqnarray}
\label{i1}
\bar \beta_0(T) &=& \frac{1}{9(1 - \alpha)^2}\frac{\Psi_0^2}{-24s\gamma\lambda}
\int_0^\infty\int_0^\infty \exp{\biggr(\frac{y}{48s\gamma}\biggl)}\exp{(-\frac{(\ln x)^2}{2\gamma})}\times \nonumber\\ 
& &\ln\biggr\{(\lambda^2 + s^2T^2)\biggr[\biggr(\arctan(sT/\lambda)\biggl)^2 + y\biggl]\biggr(- \frac{1}{48s\gamma\lambda}\biggl)\biggl\}\frac{dxdy}{x} \quad ,\\
\label{i2}
\bar \beta_+(T) &=& - \frac{1}{9(1 - \alpha)^2}\frac{\Psi_0^2}{-24\gamma s\lambda}
\int_0^\infty\int_0^\infty\exp{\biggr(\frac{y}{48s\gamma}\biggl)}\exp{(-\frac{(\ln x)^2}{2\gamma})}
\times \nonumber \\
& &\ln\biggr\{\frac{[\arctan(sT/\lambda)]^2 + y}{-96\gamma s\lambda x^{3(1 - \alpha)}}\biggl\}\frac{dydx}{x} \quad ,
\end{eqnarray}
where $x = 4^{-\frac{1}{3(1 - \alpha)}}D$ and $y = E$. The variables $x$ and $y$ were restricted to positive values in order
to assure that the metric functions are real.
Even if the integrals (\ref{i1},\ref{i2}) seem to admit no simple closed expressions, it is evident that they are time dependent.

\section{Conclusions}

It is generally expected that quantum effects in the very early universe may furnish the set of initial conditions
which will determine the subsequent evolution of the Universe when its classical phase is reached. By initial
conditions we mean here the isotropy and homogeneity. Moreover, it is also expected that those quantum effects
may lead to the avoidance of the initial singularity, one of the major problems of the standard cosmological
model. 
In this work we have tried to analyse the possibility that quantum effects can suppress
initial anisotropies. Specifically, we have studied
a Bianchi I model with a perfect fluid, with an isotropic pressure, employing the Schutz's description for
perfect fluids. This problem has for us two main interests: first, it adds more degrees of freedom
with respect to the isotropic model, since now we have four independent variables instead of just two; second,
it permits to verify if anisotropies in the early Universe disappear in the quantum model, as it happens
with the initial singularity for the corresponding isotropic one. The employement of Schutz's formalism
for the description of the perfect fluid present in the model allows us to identify quite naturally
a time coordinate associated to the matter degrees of freedom, since the canonical momentum corresponding
to the matter variables appears linearly in the Hamiltonian. Hence, the Wheeler-DeWitt equation
can be reduced to a Schr\"ordinger-like equation in terms of three dynamical
variables related to the metric function, $\beta_0$, $\beta_+$ and $\beta_-$. In order
to treat the problem analytically, we have restricted the problem to 
the special
case that the wave function is independent of one of the variables, namely $\beta_-$.
\par
The resulting Schr\"odinger-like equation has a hyperbolic signature in the kinetic term.
This means that a state of zero energy is possible along an infinite line where $p_0 = p_+$ or, equivalently, $\beta_0 = \beta_+$. This leads to an anomaly in
the boundary at infinity which, however, does not spoil the regularity of the wave function.
But as consequence, the resulting quantum system is not unitary anymore. The reason for this
loss of unitarity is the absence of a self-adjoint extension for the hermitian effective
Hamiltonian (\ref{eh}). The lack of self-adjointness in this model is due both to the
hyperbolic signature (which is a consequence of treating an anisotropic quantum cosmological
model) and due to the presence of a non-trivial measure in the original Hamiltonian (\ref{hamilton}) (which is consequence of treating a gravitational system), as a detailed inspection of the computation of the deficiency indices
reveals. The loss of unitarity leads to
a discrepancy between the many-worlds and Bohm-de Broglie interpretations of quantum
mechanics. Notice that
if we go back to the isotropic model, where there is no hyperbolicity anymore, but yet an
unusual measure, the unitarity is restored. It is important to remark that, in our opinion,
the loss of unitarity pointed out here in anisotropic models has no relation with the
description for the matter fluid, since the hiperbolicity of the kinetic term and the
unusual measure appear already in the pure gravitational sector.
\par
In order to verify this explicitly,
we have solved the Wheeler-DeWitt equation in the minisuperspace. A wave packet was constructed being regular in the sense that it is square integrable.
Using this wave packet, we have determined the behaviour of the metric functions using first the
many-worlds interpretation of quantum mechanics, which implies to compute the expectation value of those
functions. We found that there is no trace of anisotropies at any moment: the expectation value of
the function $\beta_+$ is constant while the expectation value of $\beta_0$ has essentially the
same expression as in the isotropic version of this problem. All the features of this model are the same as in the isotropic case.
Later, we have determined the behaviour of metric functions employing the ontological interpretation
of quantum mechanics, determining the bohmian trajectories. In this case the function
$\beta_+$ is no longer a constant, and an initial anisotropic Universe is predicted. Asymptotically, it becomes
isotropic like in the classical case. This result is maintained even after the averaging on the initial conditions.
As it is well known \cite{holland,nelson} the bohmian trajectories should lead to the same results that
are obtained computing the expectation values after averaging on the initial conditions. This equivalence
does not occur for the anisotropic Bianchi I cosmological model because the equivalence
between both interpretations is valid only for unitary system.
\par
Let us now precise in another way that the loss of unitarity is due to
the anomaly at the boundary at infinity generated by the hyperbolic signature of
the kinetic term. Writting the wave function as in (\ref{wf}), we obtain the expression
\begin{equation}
12e^{3(1 - \alpha)\beta_0}\frac{\partial}{\partial t}\Sigma^2 = \frac{\partial}{\partial\beta_0}
\biggr(\Sigma^2\frac{\partial}{\partial\beta_0}\Theta\biggl) - \frac{\partial}{\partial\beta_+}
\biggr(\Sigma^2\frac{\partial}{\partial\beta_+}\Theta\biggl)
\end{equation}
which should express the conservation of probability. Integrating in $\beta_0$ and $\beta_+$ does not
lead to the vanishing of the integrated left hand side, since
the norm of the wave function is time dependent. The term in the right-hand side can be converted to a surface integral at infinity. The currents are zero at infinity excepted along
the line $\beta_0 = - \beta_+$, and hence the right hand side also does not vanish.
\par
The question resulting from the analysis made above is if it is possible to consider seriously
such quantum cosmological perfect fluid model. In principle, the answer, in our opinion, must
be positive, since this system is a natural extension of the corresponding isotropic case, where everything
is well definite. But, in order to take seriously such model, the usual interpretation scheme
of quantum mechanics must be enlarged in order to take into account creation/annihilation of
probabilities and bohmian trajectories. In what concerns cosmology this perhaps may be done
since these creation/annihilation refer to disconnected universes. But, this remains an open issue,
and what we may assert at the moment is that the loss of unitarity is a general feature of
quantum system where the signature of the kinetic term is hyperbolic as it happens with anisotropic
cosmological models. Moreover, since the predictions for the evolution of the universe
using the many-worlds or the ontological interpretations do not coincide, one of them
must be more suitable for implementation of this enlarged interpretation of quantum mechanics.
Notice, finally, that the loss of unitarity appears already when quantum fields are
quantized in space-times with closed timelike curves \cite{bolware}. But this represents another context
since the space-time itself remains classical.

\vspace{0.2cm}
{\bf Acknowledgements:} We thank N. Pinto-Neto, J. Acacio de Barros and N.A. Lemos for many enlightfull
discussions and CNPq (Brazil) for partial financial support.
\appendix{
\section{Conditions for the equivalence between many-worlds and
dBB interpretations}
Let us for simplicity consider a one dimensional quantum mechanical
system:
\begin{equation}
- \frac{\partial^2}{\partial x^2}\Psi + V(x)\Psi = i\frac{\partial}{\partial t}\Psi \quad .
\end{equation}
From this expression we obtain
\begin{equation}
\frac{\partial {\cal R}}{\partial t} = - \frac{\partial}{\partial x}\biggr({\cal R}\frac{\partial}{\partial x}\theta\biggl)
\quad ,
\end{equation}
where we have used (\ref{wf}) and we have defined ${\cal R} = \Psi^*\Psi$.
The expectation value and the averaged bohmian trajectories are given by
\begin{equation}
<x>_C = \frac{\int_{-\infty}^\infty {\cal R}\,x\,dx}{\int_{-\infty}^\infty {\cal R}\,dx} \quad
\quad , \bar x_B = \int_{-\infty}^\infty {\cal R}_0x(t)\,dx_0
\end{equation}
where $C$ and $B$ stand for "Copenhagen" and "Bohm", respectively. The subscript in the
second integral indicates that the quantities must be evaluated at $t = 0$. Also in this
second integral the function $x(t)$ is obtained by integrating the bohmian
trajectories $\dot x(t) = d\Theta/dx$.
If the norm of the wave function is made equal to one in $t = 0$, those quantities are
identical at this moment.
\par
To show that they are the same for all time, we must just show that their derivatives
are the same for any value of $t$.
Taking the derivative and using the expression for the current in the first integral
and the expression for the bohmian trajectories in the second one, we obtain
\begin{equation}
\label{equiv}
\frac{d}{dt}<x>_C = \int_{-\infty}^\infty {\cal R}\;\frac{\partial\Theta}{\partial x}\;dx \quad ,
\quad \frac{d}{dt}\bar x_B = \int_{-\infty}^\infty {\cal R}_0\;\frac{\partial\Theta}{\partial x}\;dx_0
\quad ,
\end{equation}
where we have assumed the norm of the wave function equal to one.
Integrating the expression for the current, we have that the derivative of the norm
of the wave function is zero. Hence, it has the same value at any time, from which
\begin{equation}
\int_{-\infty}^\infty {\cal R}_tdx_t = \int_{-\infty}^\infty {\cal R}_0dx_0 \quad .
\end{equation}
This implies that, after a changing of variable in the second integral,
\begin{equation}
{\cal R}_0\frac{dx_0}{dx_t} = {\cal R}_t \quad .
\end{equation}
Inserting this in (\ref{equiv}) we find that the derivative of expectation value and
of the averaged bohmian trajectories are the same for any value of $t$. Since, they were equal
for $t = 0$, both quantities are identical for any value of $t$.
Notice that it was essential to have a constant norm of the wave function in order to obtain
this result. In the Bianchi I quantum model studied in this paper the analysis follows
the same lines, but the norm of the wave
function is time-dependent and the equivalence exhibited here is no longer valid.}


\begin{thebibliography}{100}
\bibitem{halliwell} J.A. Halliwell, in {\bf Quantum cosmology and baby
universes}, edited by S. Coleman, J.B. Hartle, T. Piran and S. Weinberg,
World Scientific, Singapore (1991)
\bibitem{adm} R. Arnowitt, S. Deser e C.W. Misner,{\bf Gravitation:
an introduction to current research}, edited by L. Witten, Wiley, New York
(1962);
\bibitem{isham} C.J. Isham, {\it Canonical quantum gravity and
the problem of time}, gr-qc/9304012;
\bibitem{kuchar} J.D. Brown, K.V. Kuchar, Phys. Rev. {\bf D51}, 5600(1995);
\bibitem{rubakov} V.G. Lapchinskii and V.A. Rubakov,
Theor. Math. Phys. {\bf 33}, 1076(1977);
\bibitem{gotay} M.J. Gotay and J. Demaret, Phys. Rev. {\bf D28}, 2402(1983);
\bibitem{nivaldo} N.A. Lemos, J. Math. Phys. {\bf 37}, 1449(1996).
\bibitem{flavio1} F.G. Alvarenga and N.A. Lemos, Gen. Rel. Grav.
{\bf 30}, 681(1998);
\bibitem{nelson1} J. Acacio de Barros, N. Pinto-Neto and M.A. Sagioro-Leal,
Phys. Lett. {\bf A241}, 229(1998);
\bibitem{nivaldo1} F.G. Alvarenga, J.C. Fabris, N.A. Lemos and
G.A. Monerat, Gen. Rel. Grav. {\bf 34},
651(2002);
\bibitem{brasil1} A.B. Batista, J.C. Fabris, S.V.B. Gon\c{c}alves and J. Tossa, Phys.
Rev. {\bf D65}, 063519(2002);
\bibitem{schutz1} B.F. Schutz, Phys. Rev. {\bf D2}, 2762(1970);
\bibitem{schutz2} B.F. Schutz, Phys. Rev. {\bf D4}, 3559(1971);
\bibitem{tipler} F.J. Tipler, Phys. Rep. {\bf 137}, 231(1986);
\bibitem{holland} P.R. Holland, {\bf The quantum theory of motion:
an account of the de Broglie-Bohm interpretation of quantum
mechanics}, Cambridge University Press, Cambridge(1993);
\bibitem{nelson} N. Pinto-Neto, Procedings of the VIII Brazilian School
of Cosmology and Gravitation II, Edited by M. Novello, (1999);
\bibitem{lidsey} J.E. Lidsey, Phys. Lett. {\bf B352}, 207(1995);
\bibitem{gurovich} V.N. Folomeev and V.Ts. Gurovich, Gravit\&Cosmol. {\bf 6}, 19(2000);
\bibitem{nelson2} A.F. Velasco, {\bf Cosmologia qu\^antica de teorias escalar-tensoriais na
interpreta\c{c}\~ao de Bohm-de Broglie}, PhD thesis, CBPF, Rio de Janeiro, Brazil (2000). In portuguese;
\bibitem{velasco} N. Pinto-Neto, A.F. Velasco, R. Colistete Jr., Phys. Lett. {\bf A277},
194(2000);
\bibitem{hawking} S.W. Hawking and J.C. Luttrell, Phys. Lett. {\bf B143}, 83(1984);
\bibitem{farhi} E. Farhi and S. Gutmann, Int. J. Mod. Phys. {\bf A5}, 3029(1990);
\bibitem{symon} M. Reed and B. Simon, {\bf Methods of modern mathematical physics},
Vol. II, Academic Press, New York (1975);
\bibitem{bolware} D.G. Boulware, Phys. Rev. {\bf D46}, 4421(1992).
\end{thebibliography}
\end{document}